\documentclass[aps,prd,nofootinbib,preprintnumbers,floatfix,twocolumn,superscriptaddress]{revtex4}

\usepackage{graphics}
\usepackage{graphicx}
\usepackage{tabularx}
\usepackage{makeidx}
\usepackage{epsfig}
\usepackage[caption=false]{subfig}
\usepackage{colortbl}
\usepackage{colordvi}
\usepackage{verbatim}
\usepackage{amsmath, amsthm}
\usepackage{enumerate}
\usepackage{hyperref}
\usepackage{dcolumn}
\usepackage{bm}
\usepackage{amssymb,mathrsfs}
\usepackage[normalem]{ulem}
\usepackage{tikz}
\usetikzlibrary{positioning}

\newcommand{\beq}{\begin{equation}} 
\newcommand{\eeq}{\end{equation}} 
\newcommand{\bea}{\begin{eqnarray}} 
\newcommand{\eea}{\end{eqnarray}} 
\def\benu{\begin{enumerate}}
\def\eenu{\end{enumerate}}

\def\l{\left}
\def\r{\right}
\def\d{{\rm d}}

\begin{document}
 
\title{Polymer quantization and advanced gravitational wave detector}
\author{D. Jaffino Stargen}
\email{jaffinostargen@gmail.com}
\affiliation{Chennai Mathematical Institute, Siruseri, Kelambakkam 603103, India.}
\affiliation{Department of Physics, Indian Institute of Technology Bombay, Mumbai~400076, India}
\author{S. Shankaranarayanan}
\email{shanki@phy.iitb.ac.in}
\affiliation{Department of Physics, Indian Institute of Technology Bombay, Mumbai~400076, India}
\author{Saurya Das}
\email{saurya.das@uleth.ca}
\affiliation{Department of Physics and Astronomy, University of Lethbridge,
Alberta, T1K 3M4, Canada}
\begin{abstract}
We investigate the observable consequences 
of Planck scale effects in the advanced 
gravitational wave detector by polymer 
quantizing the optical field in the 
arms of the interferometer. For large 
values of polymer energy scale, compared to the frequency of photon field in the interferometer
arms, we consider the optical field to be a collection of infinite decoupled harmonic oscillators,
and construct a new set of {\it approximated} polymer-modified 
creation and annihilation operators to 
quantize the optical field. Employing 
these {\it approximated} polymer-modified operators, we 
obtain the fluctuations in the radiation 
pressure on the end mirrors and the number 
of output photons.  We compare our 
results with the standard quantization 
scheme and corrections from the 
Generalized Uncertainty Principle.
\end{abstract}
\maketitle
\section{Introduction}
Despite various approaches to quantum gravity, a complete theory that works at Planck energies remains elusive. 
There have also been complementary approaches which attempt to build viable, self-consistent phenomenological models that look for broad features, and with robust experimental signatures~\cite{AmelinoCamelia:2004hm,Hossenfelder:2009nu,AmelinoCamelia:2008qg}. 
They capture key ingredients, such as the introduction of a new length scale, discreteness of space-time, the Generalized Uncertainty Principle (GUP) and violation of Lorentz invariance, which will remain in a complete quantum theory of gravity. 
\par
Polymer quantization is one such scheme inspired by Loop Quantum Gravity~\cite{Thiemann:2007zz,Rovelli:2004tv,Ashtekar:2002vh,Ashtekar:2002sn,Hossain:2009,Hossain:2010eb,Garcia-Chung:2016buw,Hossain:2017poa}, which captures the discreetness of the space-time 
(a key feature of all theories of quantum gravity) by introducing a fundamental scale. Due to the presence of the fundamental scale ({assumed to be of the order of Planck scale}), the Hilbert space in Polymer quantization is different from the one in canonical quantization.

\par The key distinguishing feature between the canonical quantization and Polymer quantization is the treatment of conjugate classical variables. In the case of canonical quantization of point particles in 1-dimension, Heisenberg algebra is employed; the position and momentum variables are elevated to operators and satisfy the canonical commutation relations:
\begin{equation}
[\hat{x}, \hat{x}]=0, \qquad[\hat{p}, \hat{p}]=0 \qquad[\hat{x}, \hat{p}]=i \hbar    
\end{equation}
However, in the case of polymer quantization,  the presence of a length scale makes the Weyl algebra more suited. In this case, the pair of Unitary operators ($\hat{V}, \hat{U}$) satisfy the following Weyl relations:
\begin{eqnarray}
\hat{U}(\lambda) \hat{V}(\mu) &=& 
e^{-i \hbar \lambda \mu} \hat{V}(\mu) \hat{U}(\lambda), \\
\hat{U}\left(\lambda_{1}\right) \hat{U}\left(\lambda_{2}\right)&=& \hat{U}\left(\lambda_{1}+\lambda_{2}\right)  \nonumber \\
\hat{V}\left(\lambda_{1}\right) \hat{V}\left(\lambda_{2}\right)&=&\hat{V}\left(\lambda_{1}+\lambda_{2}\right) \nonumber
\end{eqnarray}
where $\lambda$ and $\mu$ are c-numbers \cite{Takhtajan:QM}. 
The discreetness of the space-time is introduced by assuming that the quantum states are countable sums of plane waves, i.e.
\begin{equation}
\left\langle x_{i} | x_{j}\right\rangle=\delta_{i, j}  ~.
\end{equation}
Although the position operator is well-defined, 
the discreetness of geometry implies that 
the momentum operator cannot be defined. 
This will affect various physical observables, and the question which naturally arises
is whether such signatures of Planck scale effects can be measured in very-high sensitive
current and future experiments such as gravitational wave detectors. 

In the next decade, several advanced ground-based gravitational-wave detectors will be
operational with baseline up to 10 Km~\cite{CosmicExplorer,Danilishin:2019dxq}. Specifically,
the Einstein Telescope is to be built underground to reduce the seismic noise, and the Cosmic
Explorer is to use cryogenic systems to help cut down the noise experienced from the heat on its electronics. 
At low-frequency, the sensitivity of these detectors is affected by seismic and
quantum-mechanical noises (including, for example, the radiation-pressure noise). 
Thus, the advanced gravitational wave detectors may provide the unique opportunity
of distinguishing between polymer quantization and canonical quantization using the 
radiation-pressure noise 
curves.

In this work, we use the advanced LIGO configuration to obtain the fluctuations in the radiation
pressure on the {end} mirrors and that in the number of output photons in the two quantization
(polymer and canonical) schemes. More specifically, extending Caves' calculations~\cite{Caves:1981},
for small values of polymer length scale (compared to the inverse of frequency of the photon field)
we consider the field to be a collection of infinite independent harmonic oscillators, and use polymer
quantization to quantize these harmonic oscillators of the electromagnetic field in the Michelson-Morley
interferometer arms of the advanced gravitational wave detector with the advanced LIGO configuration
\footnote{Polymer quantizing the optical field, without any approximation, can modify the equation
of motion, which consequently can lead to a modification in dispersion relation. Note that we have not
taken into account the effect of this modified dispersion relation. However, we have shown in
Sec. (\ref{sec:Conclusions}) that the effects of this modified dispersion relation due to polymer
quantization on the interferometer noises are of the same order as the approximated corrections calculated in this work.}. 

The first step in the quantization of the electromagnetic field is to write the Hamiltonian as an
infinite sum of independent harmonic Oscillators. As discussed before, since we are interested
in the limit where polymer length scale $\lambda  \to 0$, it is possible to consider the optical field
as a collection of infinite number of independent oscillators. Hence this procedure is identical for
both the polymer and canonical quantization schemes. However, the difference arises in the definition
of the momentum operator in the polymer quantization. To our knowledge, the creation and annihilation
operators corresponding to the polymer quantized Harmonic oscillator has not been obtained in the literature.
In this work, for small values of polymer length scale (compared to the inverse of
frequency of the photon field), we obtain approximate creation and annihilation operators
for the individual polymer quantized harmonic oscillators. We use these operators to obtain
the fluctuations in the radiation pressure on the end mirrors, and the fluctuations in the number of output photons.

The paper is organized as follows: In Sec. (\ref{sec:Polymer1}), we briefly review polymer
quantization and its application to the simple harmonic oscillator. In
Sec. (\ref{sec:Polymer2}), we construct the approximate ladder operators
corresponding to the polymer harmonic oscillator. 
In Sec. (\ref{sec:Noise1}), we briefly review the standard analysis of 
radiation-pressure noise and photon-count noise for the advanced LIGO 
configuration~\cite{Caves:1981}. 
In Sec. (\ref{sec:Noise2}), we obtain the radiation-pressure noise and photon-count 
noise for the case of polymer quantized electromagnetic fields. Finally, in
Sec. (\ref{sec:Conclusions}), we discuss the implications of our results.

\section{Polymer Quantum Mechanics of simple harmonic oscillator}
\label{sec:Polymer1}

In this section, we briefly review polymer quantization and the polymer quantized harmonic oscillator. 
As mentioned
earlier, the polymer quantization possesses a fundamental length scale, usually assumed to
be of the order of Planck length. Thus, the structure of the Hilbert space in
polymer quantization is different from that in Canonical quantization.  
We then obtain the polymer quantized energy eigenfunctions for the harmonic oscillator. 

\subsection{Polymer Quantization}

As mentioned before, the crucial difference between the canonical and polymer
quantization procedures is the
choice of Hilbert space. The polymer Hilbert space is the space of almost periodic functions
\cite{cord:1968}, where the wave-function of a particle is expressed as the linear combination \cite{Hossain:2010eb}
\begin{equation}
 \psi({\bf p})=\sum_{j=1}^{N}c_{j} e^{i {\bf p}\cdot{\bf x}_j/ \hbar},
\end{equation}
where $\{{\bf x}_j|j=1,2,...N\}$ is a selection from ${\mathbb R}^3$. In the polymer Hilbert space, the inner product is defined as
\cite{Hossain:2010eb}:
\begin{eqnarray}
 \langle {\bf x}_i|{\bf x}_j \rangle &=& \lim_{T \to \infty} \frac{1}{(2T)^3}
 \int_{-T}^{T} \int_{-T}^{T} \int_{-T}^{T} e^{-i {\bf p}.({\bf x}_i-{\bf x}_j)/\hbar} \d^3 p
 \nonumber \\
 &=&\delta_{i,j}.
\end{eqnarray}
Note that the plane waves are normalizable in polymer Hilbert space.

The configuration and translation operators in the polymer Hilbert space are \cite{Hossain:2010eb}
\begin{equation}
{\hat {\bf x}}=i\hbar \nabla_{\bf p},~~~~~~ {\hat U}_{\lambda}={\widehat {e^{i\lambda {\bf p}}/\hbar}},
\end{equation}
which act as
\begin{equation}
\label{polymer-operators}
 {\hat {\bf x}} e^{i {\bf p}\cdot{\bf x}_j/\hbar}={\bf x}_j e^{i {\bf p}\cdot{\bf x}_j},~~~~
{\hat U}_{\lambda} e^{i {\bf p}\cdot{\bf x}_j/\hbar}=e^{i {\bf p}\cdot({\bf x}_j+{\boldsymbol \lambda})/\hbar},
\end{equation}
where $\lambda$ is the fundamental (polymer) length scale. 
Due to the discreteness of the geometry the momentum operator is not well defined~\cite{Halvorson:2004, Hossain:2010eb}, while the position operator is well defined. However, an effective momentum operator can be defined as\cite{Hossain:2010eb}: 
\begin{equation}
\label{def:PolyMomen}
\hat{P}_{\lambda} \equiv \frac{\hbar}{2 i \lambda}\left(\hat{U}_{\lambda} -\hat{U}^{\dagger}_{\lambda}\right) \, .
\end{equation}
In the limit, $\lambda \to 0$, the above definition of the effective momentum operator leads to the momentum operator in canonical quantization. The explicit dependence of momentum operator on the fundamental length scale $\lambda$ is the key feature of polymer quantization, which leads to the effects of polymer quantization on a given system. In other words, the classical observables which depend on momentum become $\lambda$-dependent operators in polymer quantum mechanics. For example, the polymer Hamiltonian
operator corresponding to the classical Hamiltonian 
$H=p^2/2m+V(x)$ is
\begin{equation}
{\hat H}_{\lambda}=\frac{{\hat P}_{\lambda}^2}{2m}+V({\hat x}),
\end{equation}
where $m$ is the mass of the particle, and $V(x)$ is the external potential. The effect of Polymer quantization 
on the energy eigenvalues enter through ${\hat P}_{\lambda}$.

\subsection{Polymer quantized simple harmonic oscillator}
The Hamiltonian corresponding to the harmonic oscillator (with frequency $\omega$) 
in the polymer quantization is:
\begin{equation}
 {\hat H}_{\lambda}=\frac{{\hat P}_{\lambda}^2}{2m}+\frac{1}{2}m\omega^2 {\hat x}^2.
\end{equation}
Using (\ref{def:PolyMomen}) in the above Hamiltonian, in momentum basis,
the energy eigenvalue equation ${\hat H}_{\lambda}\Psi(p)=E\Psi(p)$ becomes~\cite{Hossain:2010eb}:
\begin{equation}
\label{Eqn:energy-eigen}
 \frac{\d^2 \Psi(z)}{\d z^2}+\l[\alpha-2q {\rm cos}(2z)\r]\Psi(z)=0 \, ,
\end{equation}
where,
\begin{eqnarray}
\label{eq:variable-defs-1}
z&=&\frac{\lambda p}{\hbar}-\frac{\pi}{2},~~
\alpha=\frac{2E}{\hbar\omega\beta^2}-\frac{1}{2\beta^4},\\
\label{eq:variable-defs-2}
q&=&\frac{1}{4\beta^4},~~~~~~
\beta=\lambda \sqrt{\frac{m\omega}{\hbar}}.
\end{eqnarray}
Note that Eq.(\ref{Eqn:energy-eigen}) is the well-known Mathieu differential equation \cite{Abramowitz}, which 
has periodic solutions for special values of $\alpha$, i.e.,
\begin{eqnarray} 
\label{eq:PolyPsi1}
\Psi_{2n}(z) = \frac{\l(2\beta/\pi\r)^{1/2}}{\l(\hbar m\omega\r)^{1/4}}
 {\rm Ce}_{n}\l(q,z\r) &{\rm for}& \alpha=A_n(q),~~\\
\label{eq:PolyPsi2}
 \Psi_{2n+1}(z)= \frac{\l(2\beta/\pi\r)^{1/2}}{\l(\hbar m\omega\r)^{1/4}}
 {\rm Se}_{n+1}\l(q,z\r) &{\rm for}& \alpha=B_n(q)~~
\end{eqnarray}
where $A_n$, $B_n$ are Mathieu characteristic values, and
${\rm Ce}_{n}$, ${\rm Se}_{n}$ are Mathieu functions. These
functions  are $\pi$-periodic {(functions with a periodicity $\pi$)}
for even $n$ and $\pi$-antiperiodic for odd $n$~\cite{Abramowitz}.
It is important to note that these solutions are normalizable,
and in the limit $\lambda \to 0$ we recover the standard quantization mode
functions. We show this explicitly in the next section
(See Eqs. (\ref{asymptotic-1}) and (\ref{asymptotic-2})).

\begin{figure}[!t]
\begin{center}
\includegraphics[width=8.50cm]{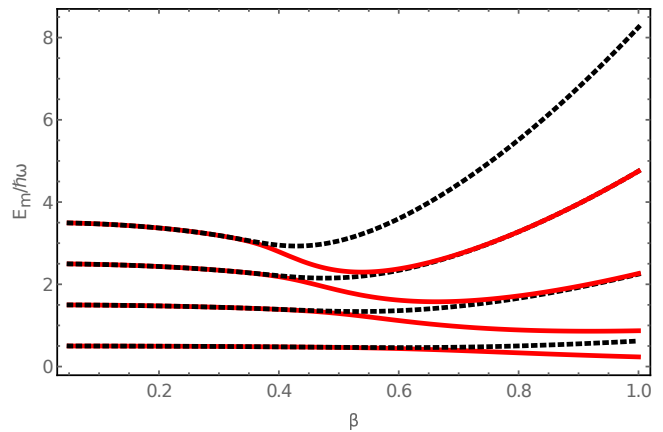} 
\caption{Plot of $E_m/\hbar\omega$ of polymer harmonic oscillator as a function
of $\beta$ as obtained in Ref.~\cite{Hossain:2010eb}.
The red and black (dotted) curves correspond to even ($m = 2 n$) and odd ($m = 2n + 1$) energy
levels, respectively. The energy levels are degenerate up to a critical value of $\beta$.}
\label{fig:energy}
\end{center}
\end{figure}

The energy eigenvalues for the even and odd quantum numbers are
{\small
\begin{equation}
 \frac{E_{2n}}{\hbar\omega}=\frac{2\beta^4 A_n(q)+1}{4\beta^2}~;~
 \frac{E_{2n+1}}{\hbar\omega}=\frac{2\beta^4B_{n+1}(q)+1}{4\beta^2}.
 \label{eq:PolyEnergy}
\end{equation}
}
In the limit of $\beta \to 0~(q \to \infty)$, the above energy eigenvalues
smoothly go over to the standard harmonic oscillator energy eigenvalues
(see Eq.(\ref{polymer-beta})). Energy levels of polymer harmonic oscillator,
$E_{2n}$ and $E_{2n + 1}$, are degenerate up to a critical value of $\beta$.
Furthermore, as $\beta$ is increased, 
the energy levels dip below (rise above) the
$\beta=0$ value for even (odd) $n$,
see Fig. (\ref{fig:energy}).
By contrast, in the case of the GUP, the 
energy levels increase monotonically above
the standard energy levels, for every $n$
\cite{kmm}.

\section{Construction of  approximated polymer ladder operators}
\label{sec:Polymer2}

We aim to obtain the fluctuations in the radiation pressure on the
end mirrors and that in the number of output photons in the 
advanced gravitational wave interferometers in the two quantization (canonical 
and polymer) schemes.

\begin{widetext}
\begin{figure*}[!htb]
\begin{center}
\begin{tabbing}
\includegraphics[width=8.50cm]{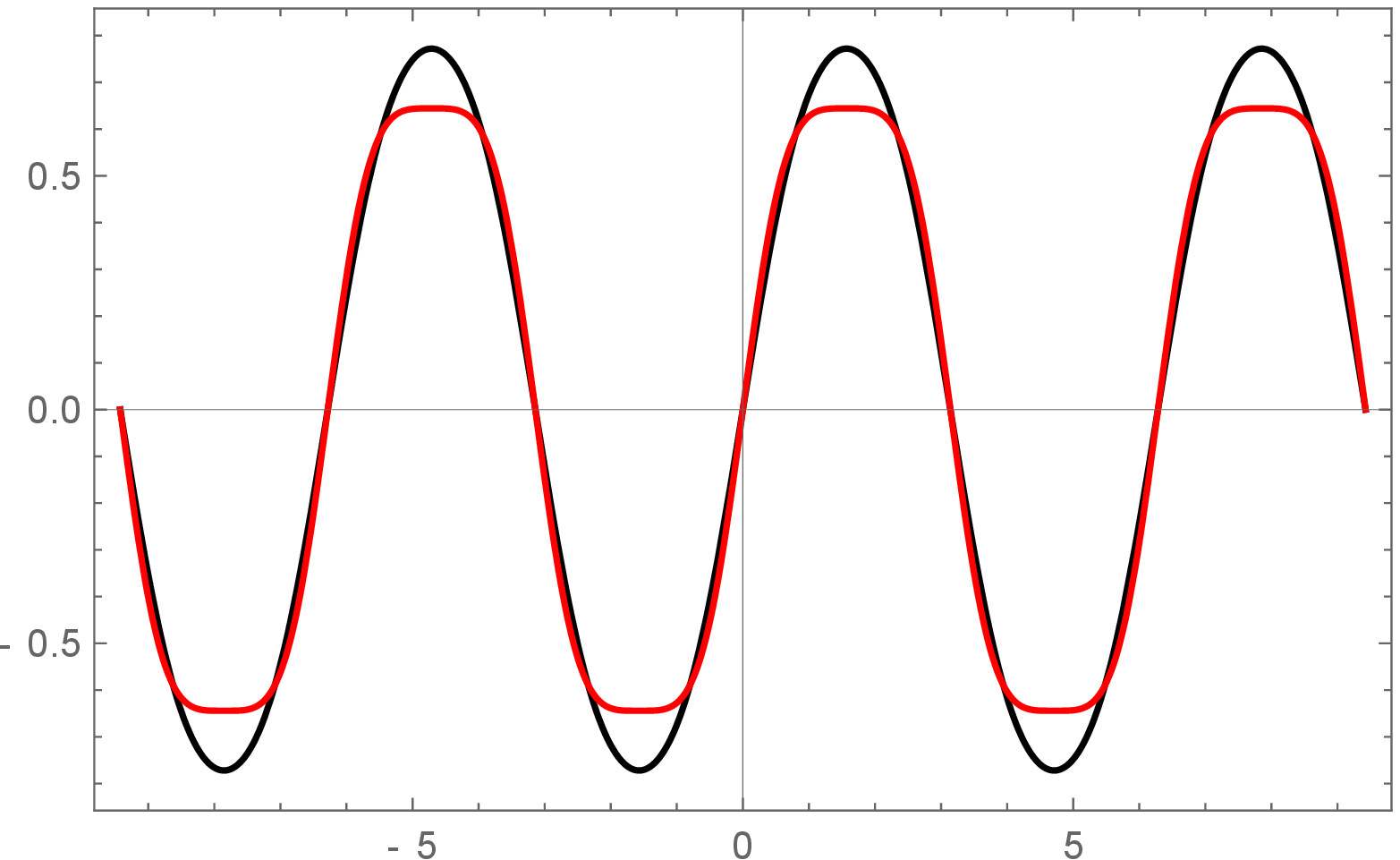}
\=\includegraphics[width=8.50cm]{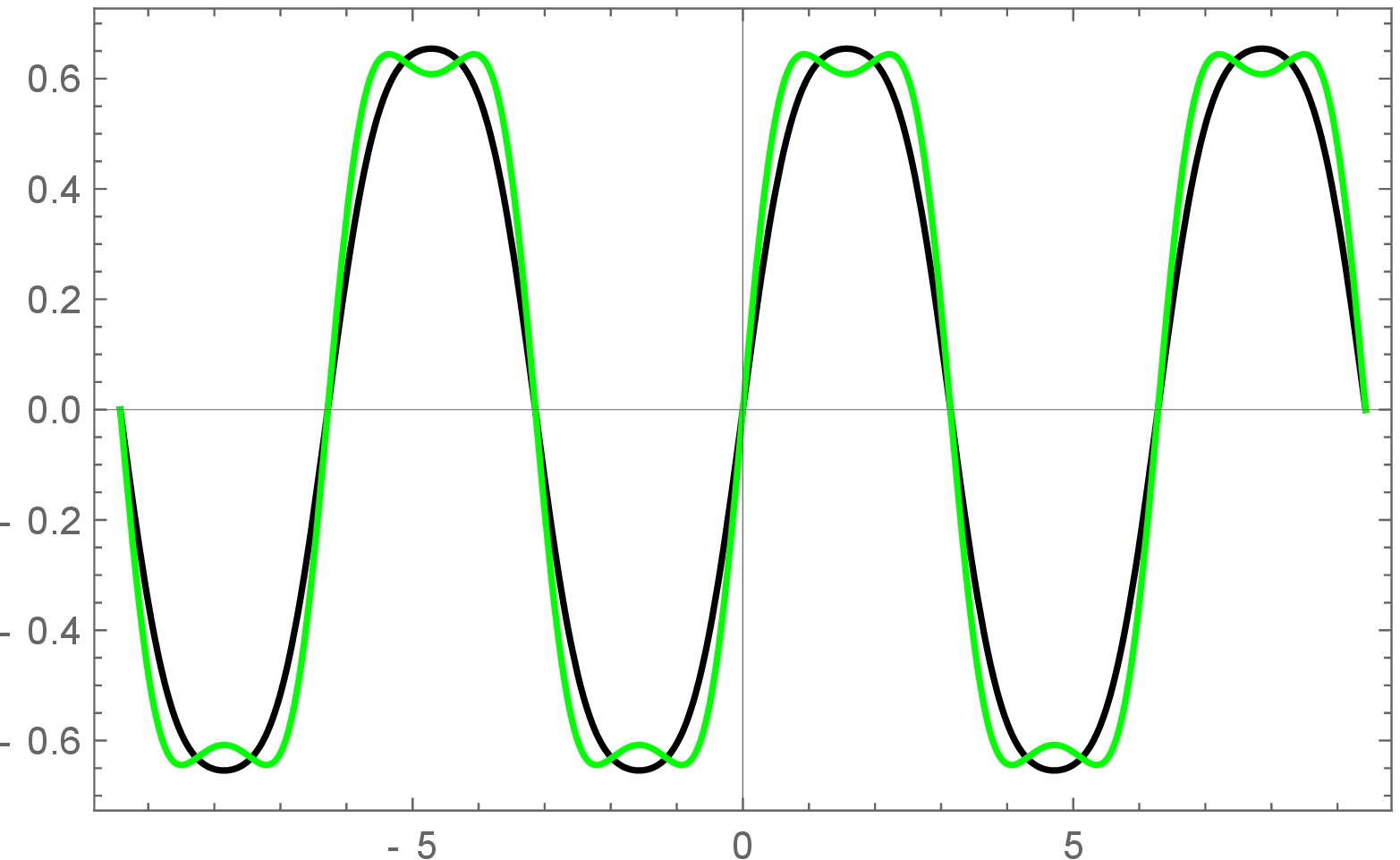}\\
\includegraphics[width=8.50cm]{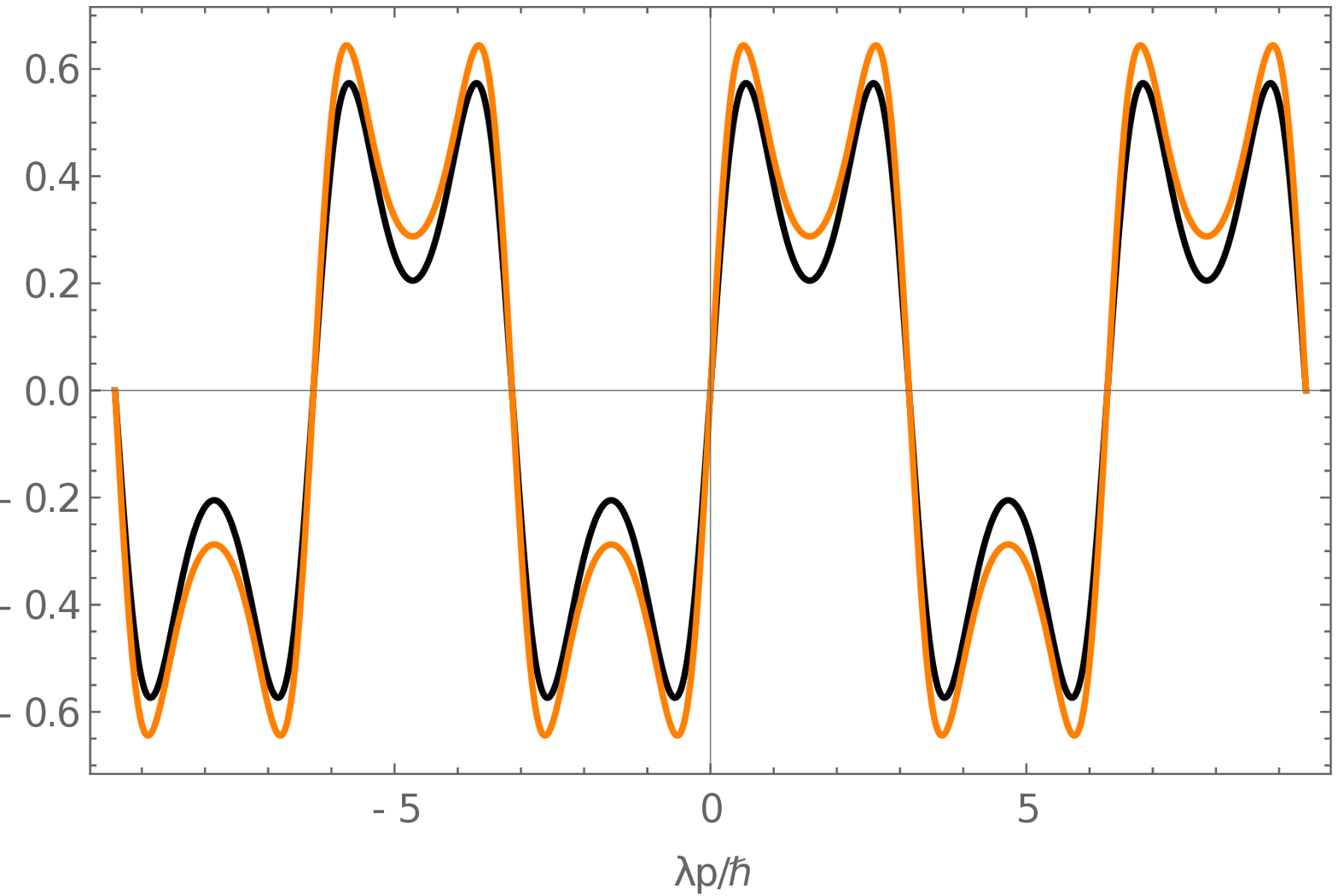}\>
\includegraphics[width=8.50cm]{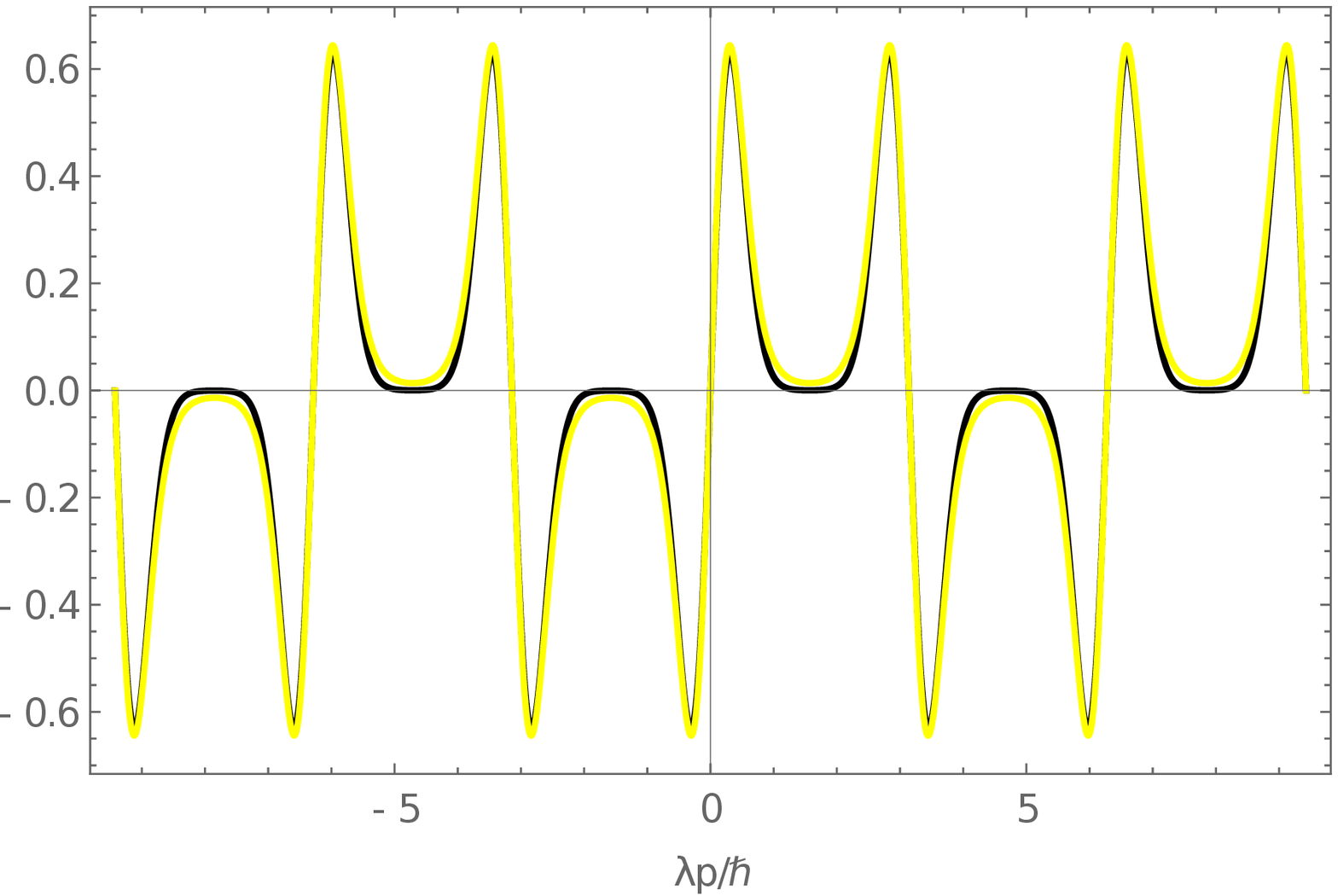}\\
\end{tabbing}
\caption{The plot of the scaled Polymer quantized Harmonic oscillator eigenfunctions 
versus dimensionless momentum ${\tilde p}=\lambda p/\hbar$. The black curve in all the
four plots above correspond to the exact Polymer eigenfunction $(\hbar m\omega)^{1/4}
\, \Psi_{ n= 2}(\tilde{p})$ as in Eq. (\ref{eq:PolyPsi1}). The Red, Green, Orange and
Yellow curves correspond to the  approximated Polymer eigenfunction
$(\hbar m\omega)^{1/4}~{\widetilde \Psi}_{n = 2}({\tilde p})$ for $\beta=1$ (red),
$0.8$ (Green), $0.5$ (Orange), and $0.3$ (Yellow), 
respectively.}
\label{fig:poly-compare-2} 
\end{center}
\end{figure*}
\end{widetext}

To compare these quantum-noises in the two quantization schemes, we need
to obtain the equivalent set of {\it approximated} creation and annihilation operators
in the polymer quantization scheme.
Obtaining the ladder operators for the polymer harmonic oscillator
is non-trivial for the following reasons: First, in standard
quantization, a linear combination of position (${\hat x}$)
and momentum (${\hat p}$) operators
can raise or lower an energy eigenstate. In the case of polymer
quantization, as can be seen from the momentum definition Eq. (\ref{def:PolyMomen}),
a simple linear combination of ${\hat x}$ and ${\hat P}_{\lambda}$
does not lead to ladder operators, which can raise or lower a given polymer energy
eigenstate. Second, from Fig. \ref{fig:energy} it is evident that
for $\beta \to 1$, the polymer energy eigenvalues Eq. (\ref{eq:PolyEnergy})
are degenerate~\cite{Hossain:2010eb}. Therefore, to construct the ladder
operators for the case of polymer harmonic oscillator,
we adopt the following procedure: we define
$|0\rangle$ to be the ground state of the polymer
harmonic oscillator. Let ${\hat {\alpha}}_\lambda$ 
be the annihilation operator satisfying the condition
${\hat {\alpha}}_\lambda|0\rangle=0$. 
As mentioned earlier, it is non-trivial to construct an exact annihilation operator 
$\hat{\alpha}_\lambda$ satisfying the above condition, especially in the limit of 
$\beta \to 1~(q \to 1/4)$. In the rest of this section, we construct the approximate 
creation and annihilation operators $(\hat{A}_{\lambda}, \hat{A}_{\lambda}^\dagger)$
for $\beta\ll1$, satisfy the condition:
${\hat A}_\lambda |{\tilde 0}\rangle=0$,
where $|{\tilde 0}\rangle$ is the approximate ground state  of the 
Polymer harmonic oscillator valid in the limit $\beta \ll 1$. 
For small values of $\beta$, the polymer energy eigenfunctions (Eqs. (\ref{eq:PolyPsi1}) and (\ref{eq:PolyPsi2})) 
can be expanded as (see Appendix. \ref{Appendix-A}) 
\begin{widetext}
\begin{eqnarray}
\label{asymptotic-1}
 \Psi_{2n}(p)=\l(\frac{1}{\pi\hbar m\omega}\r)^{1/4}
 \frac{e^{-\alpha^2/2}}{2^{n/2}\sqrt{n!}} \biggl\{{\rm H}_n(\alpha)
 &-&\frac{\beta^2}{4}\biggl[\frac{1}{32}H_{n+4}(\alpha)
 +\frac{1}{4}H_{n+2}(\alpha)+\frac{2n+1}{2}H_{n}(\alpha) \nonumber\\
 &+&n(n-1)H_{n-2}(\alpha)
 -12~nC_4 H_{n-4}(\alpha)\biggr]+O(\beta^4)\biggr\},
\end{eqnarray}

\begin{eqnarray}
\label{asymptotic-2}
 \Psi_{2n+1}(p)=\l(\frac{1}{\pi\hbar m\omega}\r)^{1/4}
 \frac{e^{-\alpha^2/2}}{2^{n/2}\sqrt{n!}} \biggl\{{\rm H}_n(\alpha)
 &-&\frac{\beta^2}{4}\biggl[\frac{1}{32}H_{n+4}(\alpha)
 -\frac{1}{4}H_{n+2}(\alpha)-\l(\frac{2n+1}{2}-\alpha^2\r)H_{n}(\alpha) \nonumber\\
 &-&n(n-1)H_{n-2}(\alpha)
 -12~nC_4 H_{n-4}(\alpha)\biggr]+O(\beta^4)\biggr\},
\end{eqnarray}
\end{widetext}
\begin{widetext}
\begin{figure*}[!htb]
\begin{center}
\begin{tabbing}
\includegraphics[width=8.50cm]{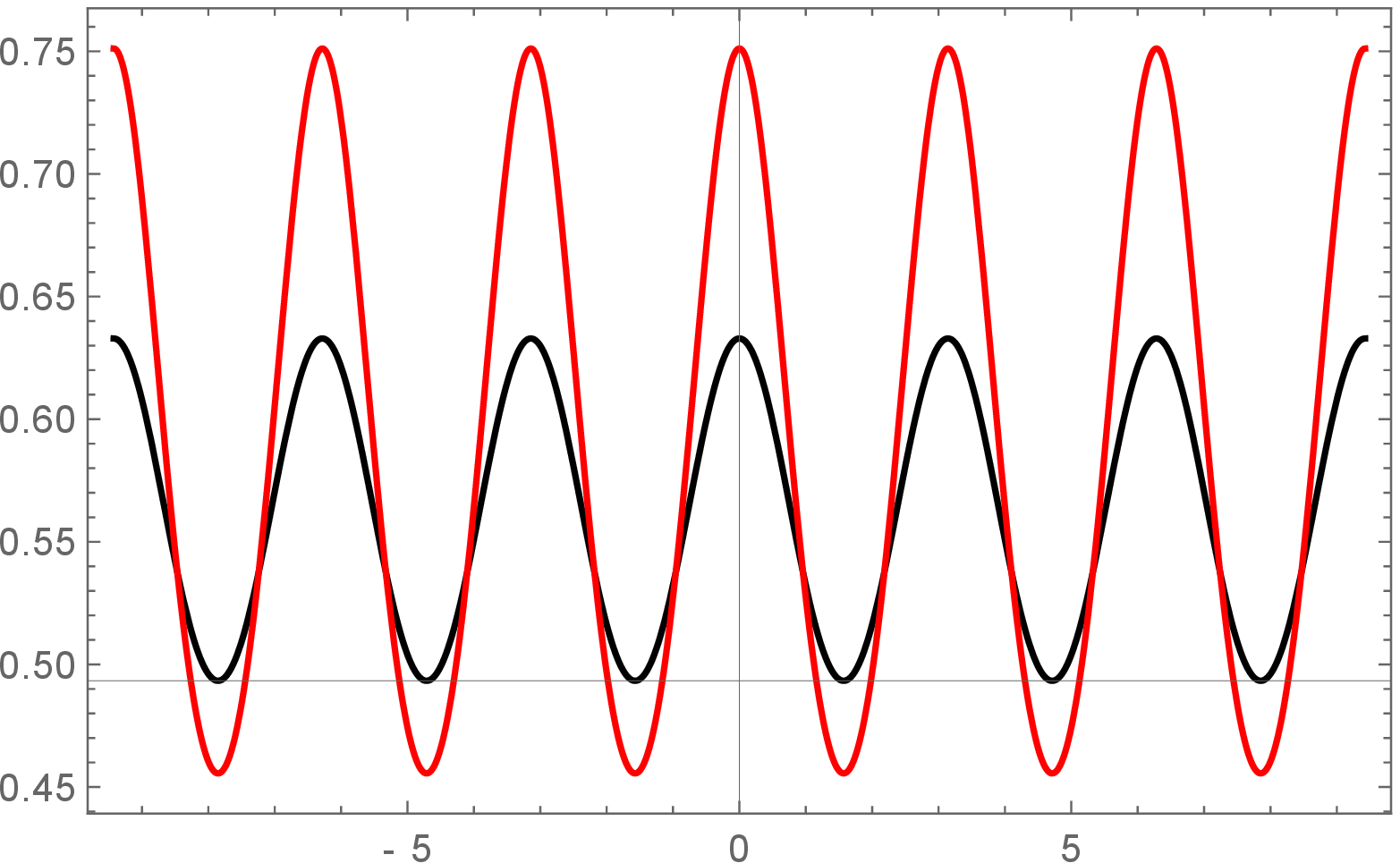}
\=\includegraphics[width=8.50cm]{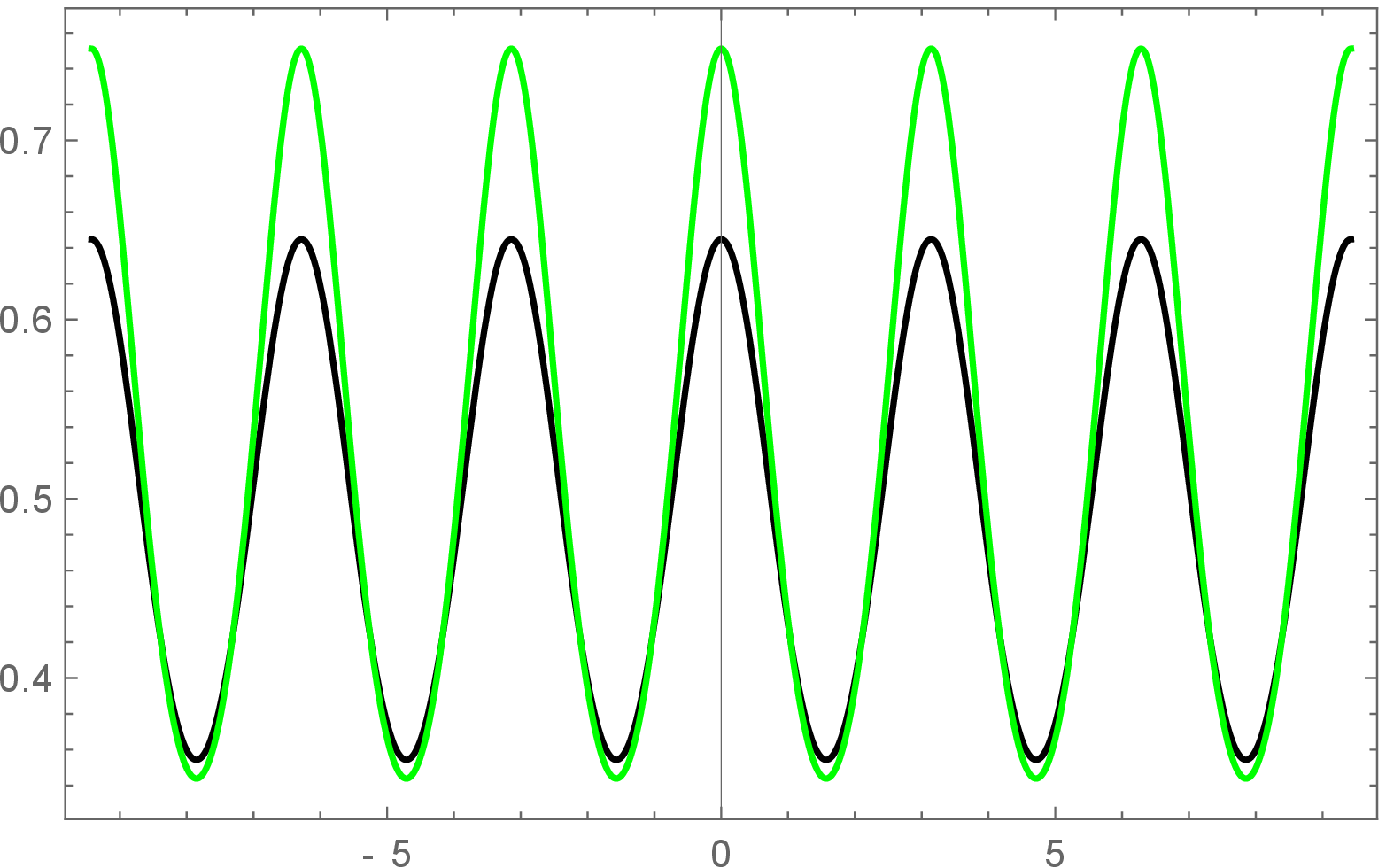}\\
\includegraphics[width=8.50cm]{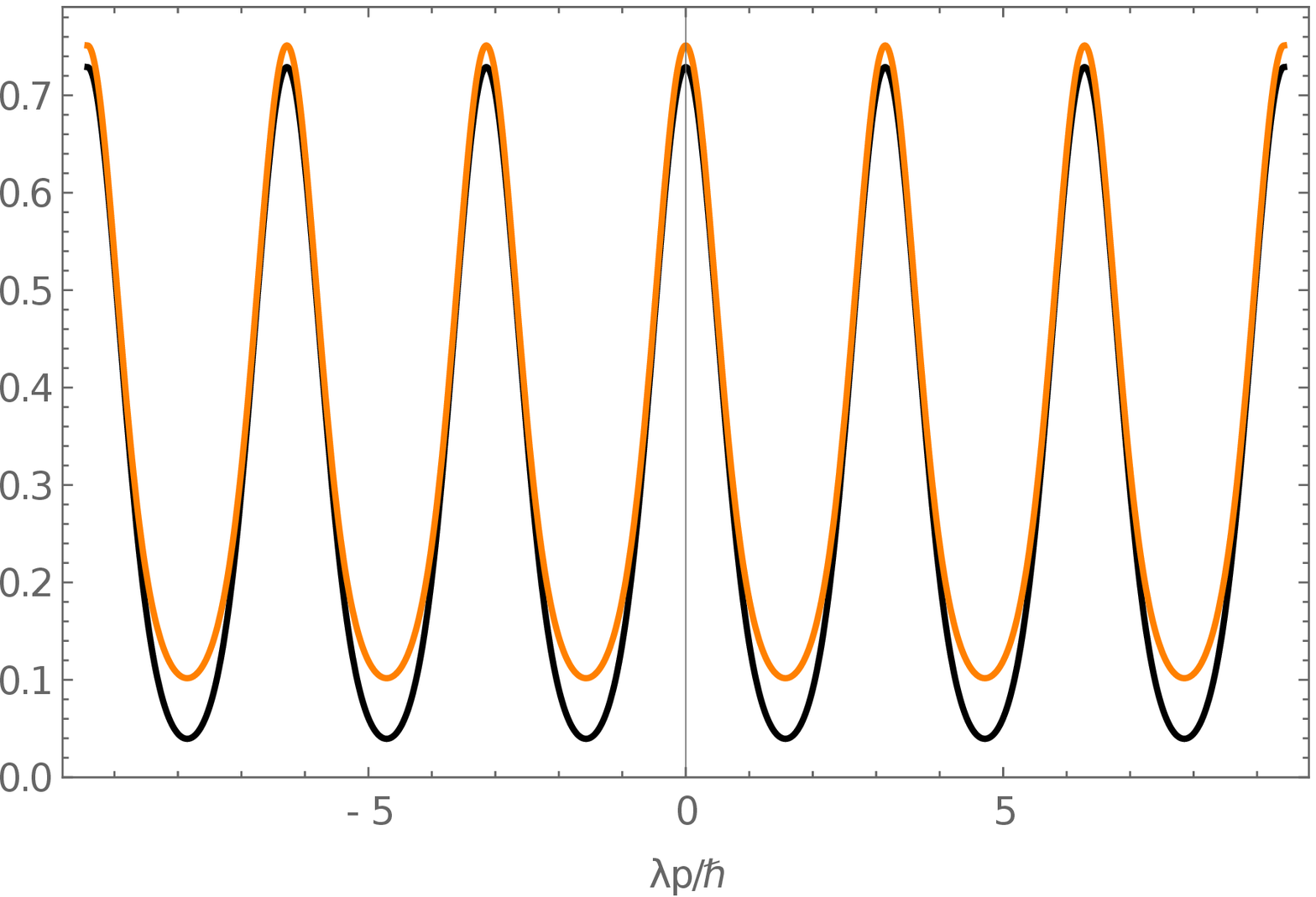}\>
\includegraphics[width=8.50cm]{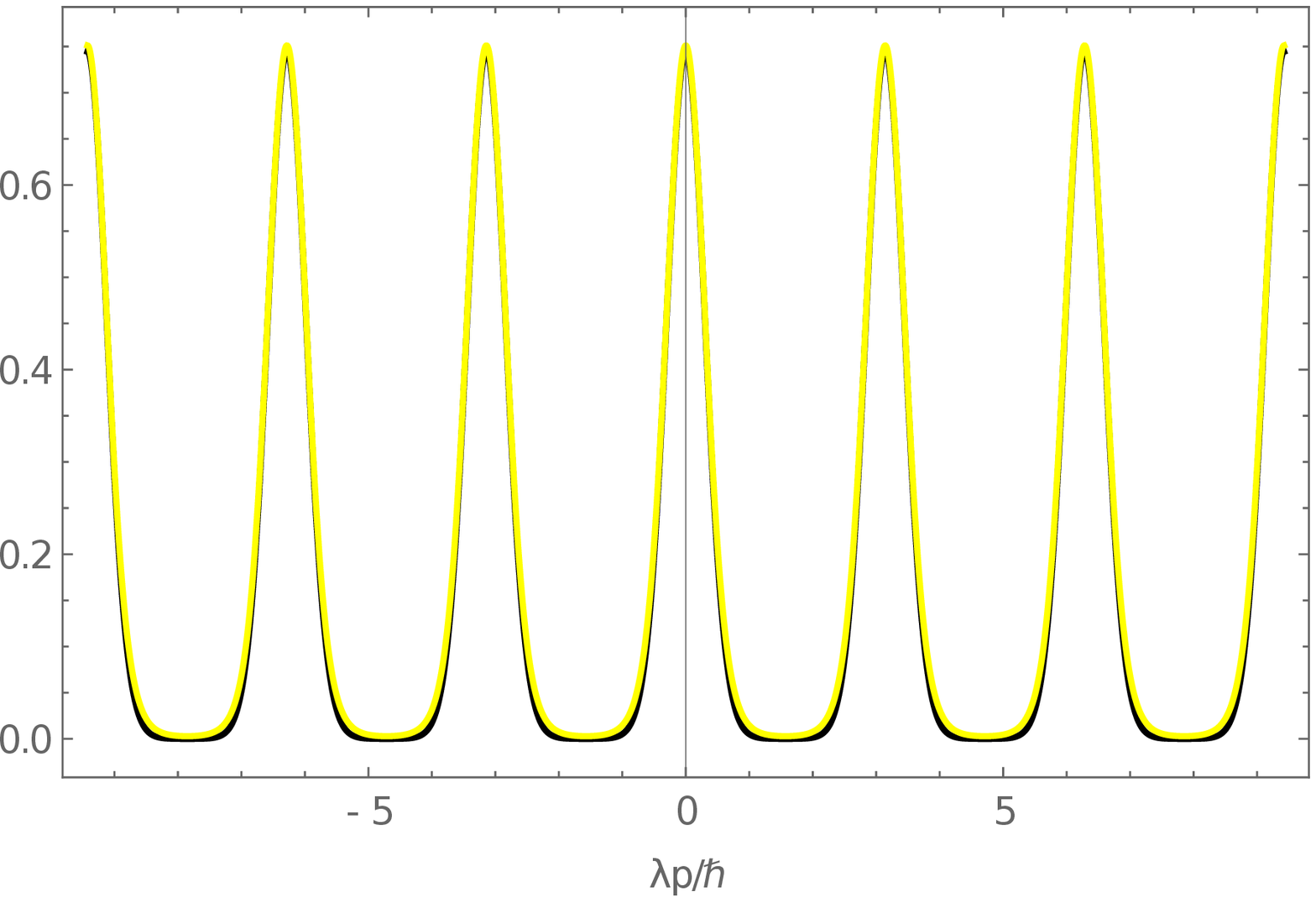}\\
\end{tabbing}
\caption{The plot of the scaled Polymer quantized Harmonic oscillator eigenfunctions 
versus dimensionless momentum ${\tilde p}=\lambda p/\hbar$. The black curve in all
the four plots above correspond to the exact Polymer eigenfunction
$(\hbar m\omega)^{1/4} \, \Psi_{ n= 0}(\tilde{p})$ as in Eq. (\ref{eq:PolyPsi1}).
The Red, Green, Orange and Yellow curves correspond to the  approximated Polymer
eigenfunction $(\hbar m\omega)^{1/4}~{\widetilde \Psi}_{n = 0}({\tilde p})$
for $\beta=1$ (red), $0.8$ (Green), $0.5$ (Orange), and $0.3$ (Yellow), respectively.}
\label{fig:poly-compare-1} 
\end{center}
\end{figure*}
\end{widetext}
where $\alpha={\rm sin}(\lambda p/\hbar)/\beta$, and $H_{n}$
are the Hermite polynomials. In the leading order in $\beta$,
we can approximate the polymer harmonic oscillator
energy eigenfunctions (Eqs. (\ref{eq:PolyPsi1}) and (\ref{eq:PolyPsi2})) as 
\begin{equation}
\label{eq:PolyPsiApprox}
 \Psi_{2n}(p)\approx\widetilde{\Psi}_{2n}(p)=
 \frac{1}{\l(\pi\hbar m\omega\r)^{1/4}}
 \frac{e^{-\alpha^2/2} {\rm H}_n(\alpha)}{2^{n/2}\sqrt{n!}} \, .
\end{equation}
Since, as mentioned earlier,
  $\Psi_{2n}(p)$ and $\Psi_{2n + 1}(p)$ are degenerate for $\beta\ll1$,
  considering only the even (or odd) eigenfunctions is
  adequate.
Figs. \ref{fig:poly-compare-1} and \ref{fig:poly-compare-2} contain the 
plots of the exact and approximate polymer harmonic oscillator eigenfunctions.
As can be seen from the figures, the approximate eigenfunctions is an excellent
approximation of the exact eigenfunctions for small values of $\beta$,
i.e., for $\beta\ll1$. Using the approximate polymer harmonic oscillator eigenfunctions, 
we can now construct the approximate annihilation operator (${\hat A}_\lambda$) in
the momentum basis.
Using the properties of Hermite polynomials, 
we can rewrite the approximate eigenfunctions (Eq. \ref{eq:PolyPsiApprox}) as:
\begin{equation}
\widetilde{\Psi}_{n}(p)= 
\frac{1}{\l(\pi\hbar m\omega\r)^{1/4}} 
\frac{1}{\sqrt{n!}} 
\l[\frac{1}{\sqrt{2}} \l(\alpha-\frac{\d}{\d\alpha} \r)\r]^n
e^{-\alpha^2/2}.
\end{equation}
As in canonical quantization, we can then write the approximate
creation operator in the polymer quantization as:
\begin{equation}
{\hat A^\dagger}_\lambda=\frac{1}{\sqrt{2}} \l(\alpha-\frac{\d}{\d\alpha} \r).
\end{equation}
Thus, the approximate ladder operators corresponding to the
polymer quantized harmonic oscillator energy eigenstates are:
\begin{eqnarray}
 {\hat A}_\lambda &=& \frac{1}{(2\hbar m\omega)^{1/2}}
 \l({\hat P}_{\lambda}-i\frac{m\omega {\hat x}}{{\rm cos}(\lambda p/\hbar)}\r) \, , \\
 {\hat A^\dagger}_\lambda&=&\frac{1}{(2\hbar m\omega)^{1/2}}
 \l({\hat P}_{\lambda}+i\frac{m\omega {\hat x}}{{\rm cos}(\lambda p/\hbar)}\r),
\end{eqnarray}
It is easy to verify that the approximate ladder operators, corresponding
to the  approximate ground state of polymer harmonic oscillator
(Eq. \ref{eq:PolyPsiApprox}),
satisfy the commutation relation $[{\hat A}_\lambda,{\hat A^\dagger}_\lambda]=1$
and ${\hat A}_\lambda |{\tilde 0}\rangle=0$.
Expanding the energy eigenvalues (Eq. \ref{eq:PolyEnergy})
for small values of $\beta$ leads to:
\begin{equation}
\label{polymer-beta}
 \frac{E_{2n}}{\hbar\omega}\approx 
 \l(n+\frac{1}{2}\r)-\frac{\beta^2}{16}\l[(2n+1)^2+1\r]+O(\beta^4) \, .
\end{equation}
Retaining only the leading order term allows us to interpret
that the $n^{\rm th}$ energy level $|{\widetilde n} \rangle$
contains $n$ particles, i.e.,
${\hat A}^\dagger_\lambda 
{\hat A}_\lambda |{\widetilde n} \rangle =n |{\widetilde n} \rangle$. 
In the rest of this work, we will use the approximate ladder
operators and energy values to study the implications of polymer
quantization on various noises in advanced LIGO configuration.
\section{Quantum-mechanical noises in advanced LIGO: Canonical Quantization}
\label{sec:Noise1}
In this section, we briefly review 
Caves' analysis for the advanced LIGO configuration~\cite{Caves:1981}.
As it is well known a gravitational wave detector is a two-arm, multi-reflecting,
laser powered Michelson--Morley interferometer \cite{Caves:1981}.
The interferometer measures the spatial strain produced by gravitational
waves as a variation in lengths of its mutually perpendicular arms,
with end mirrors attached to it. The accurate measurement of this
spatial strain is limited by two main sources
of quantum-mechanical noise --- fluctuations in the radiation pressure
on the end mirrors (radiation-pressure noise) and fluctuations in the
number of output photons (photon-count noise).

Radiation pressure noise
is due to the transfer of momentum, possessed by the optical field in the interferometer
arms, to the end mirrors. On the other hand, the photon-counting error is due
to the uncertainty produced by the photo-detectors capturing the photons leaving the
interferometer arms. As mentioned in the introduction, at low-frequency,
the sensitivity of these detectors is affected by seismic and
radiation-pressure noise. Thus, the Einstein telescope
will be sensitive to the radiation pressure. Thus,
the advanced gravitational wave detectors provide a unique opportunity
to distinguish between polymer quantization and canonical quantization using
the radiation pressure noise curves.

\subsection{Radiation-pressure noise}

As mentioned above, the radiation-pressure noise is due to the transfer of radiation
field's momentum to the end mirrors. Therefore, radiation-pressure noise
is calculated by estimating the momentum carried by the radiation field in the
arms of the interferometer. To carry out this, one requires four modes of the electromagnetic radiation field. Two modes, referred to as $E_1^+$ and $E_2^+$, are the in-modes, and the remaining two modes ($E_1^-$ and $E_2^-$) are out-modes . Among the in-modes, $E_1^+$ mode corresponds to the radiation field of frequency $\omega$ from the input
laser port, and the $E_1^-$ mode describes radiation field from the ``unused" port.

After the in-modes are scattered by the beam splitter, the ``in" and ``out" modes are related by~\cite{Caves:1981}:
\begin{eqnarray}
 E^-_{1}=\frac{e^{-i\Delta}}{\sqrt{2}} \l(E^+_{1}+e^{-i \mu} E^+_{2}\r),\\
 E^-_{2}=\frac{e^{-i\Delta}}{\sqrt{2}} \l(E^+_{2}-e^{i \mu} E^+_{1}\r),
\end{eqnarray}
where $\Delta$ and $\mu$ are the overall phase shift and relative phase shift, respectively. They depend on the intrinsic properties of the beam splitter.

The creation and annihilation operators corresponding to the input and the output modes of the interferometer are related by 
\begin{eqnarray}
 \hat{b}_{1}=\frac{e^{i\Delta}}{\sqrt{2}} \l(\hat{a}_{1}+e^{i \mu} \hat{a}_{2}\r);\\
 \hat{b}_{2}=\frac{e^{i\Delta}}{\sqrt{2}} \l(\hat{a}_{2}-e^{-i \mu} \hat{a}_{1}\r),
\end{eqnarray}
where $\hat{a}_{1}$, $\hat{a}_{2}$ are the annihilation operators corresponding
to the input ports $(1^+)$ and $(2^+)$, and $\hat{b}_{1}$, $\hat{b}_{2}$
denote the annihilation operators corresponding to the output ports $(1^-)$ and $(2^-)$
of the interferometer.

The difference between the momenta transferred to the end masses is given by~\cite{Caves:1981}:
\begin{eqnarray}
 {\hat {\cal P}}&\equiv& \frac{2 \nu \hbar \omega}{c} \l(\hat{b}_{2}^\dagger\hat{b}_{2}
 -\hat{b}_{1}^\dagger\hat{b}_{1}\r) \nonumber \\
 &=&- \frac{2 \nu \hbar\omega}{c} \l(e^{i\mu} \hat{a}_{1}^\dagger\hat{a}_{2}
 +e^{-i\mu} \hat{a}_{2}^\dagger\hat{a}_{1}\r).
\end{eqnarray}
where $\nu$ is the number of times the photon bounces in the interferometer before it reaches the receiver. 
Squeezed states are useful for the detection of the gravitational waves, as they have a reduced uncertainty in one component of the complex amplitude. The {squeezed} state of the 
electromagnetic field can be expressed as~\cite{Caves:1981}:
\begin{equation}
\label{def:Sqstate}
|\psi\rangle=S_2(\xi)D_1(\alpha) |0\rangle   \quad   \xi=-r e^{i\theta},
\end{equation}
{where $D_1(\alpha) \equiv e^{-|\alpha|^2/2} ~ e^{\alpha a^\dagger_1}$
is the displacement operator corresponding to the mode $E_1^+$, and $\alpha$
as a complex number. The squeezing operator corresponding to the mode $E_2^+$
is defined as $S_2(\xi) \equiv 
{\rm exp} \{[\xi^* a_2^2-\xi (a_2^\dagger)^2]/2\}$, with $\xi=-r e^{i\theta}$.
Note that $r$ and $\alpha$ are the squeezing parameters, and $\theta$
is referred to as squeezing angle.}

{The squeezed states satisfy the following relations:}
\begin{eqnarray}
& &  \langle \psi | \hat{\cal P} |\psi\rangle =0,\\
& & \langle \psi | \l( \Delta{\cal P} \r)^2 |\psi\rangle =
 (2 \nu \hbar\omega/c)^2 \biggl[|\alpha|^2 {\rm cosh}(2r)+{\rm sinh}^2r 
 \nonumber \\
 &+&\l(\alpha^2e^{i(\theta+2\mu)}+\alpha^{*^2}e^{-i(\theta+2\mu)}\r)~
 {\rm sinh}r~{\rm cosh}r\biggr].~~~ 
\end{eqnarray}
If $\alpha$ is real, and the squeezing angle is chosen to be $\theta=-2\mu$,
then
\begin{equation}
 \langle \psi | \l( \Delta{\cal P} \r)^2 |\psi\rangle=
 (2 \nu \omega)^2 (\alpha^2 {\rm e}^{2r}+{\rm sinh}^2r).
\end{equation}
Note that $\mu$ is characteristic of beam splitter, hence, we can always choose $\mu$ to have a particular value. 

Thus, the difference in momentum transfer on the end mirrors, for a duration of time $\tau$, leads to an error in the difference in position  of the two mirrors, $z=z_2-z_1$, is given by:
\begin{equation}
\label{eq:StandRP}
 \l(\Delta z\r)_{\rm rp}= \frac{\nu \hbar\omega\tau}{mc} \l(\alpha^2 {\rm e}^{2r}+{\rm sinh}^2r\r)^{1/2}.
\end{equation}
where $z_1$ and $z_2$ are the positions of the end mirrors.
Note that the uncertainty in the radiation pressure translates to the error
in the measurement of the gravitational wave signal and it depends on
the parameters of the radiation field in the arms of the interferometer
and the duration of time~$\tau$.

\subsection{Photon-count noise}
Photon-count error is due to the fluctuations in number of photons
leaving the arms of the interferometer. The ``in" and ``out" modes are
related by
\begin{eqnarray}
 E^-_{1}=e^{-i\Phi} \l[E^+_{2} {\rm cos}(\phi/2)
 +ie^{i \mu} E^+_{1} {\rm sin}(\phi/2)\r],\\
 E^-_{1}=e^{-i\Phi} \l[E^+_{1} {\rm cos}(\phi/2)
 +ie^{-i \mu} E^+_{2} {\rm sin}(\phi/2)\r],
\end{eqnarray}
where $\phi$ is the phase difference between the output light emitted
by the interferometer arms, and $\Phi$ is the mean phase. They are
related to the positions of the end mirrors and the parameters of the
beam splitter by the following relations \cite{Caves:1981}:
\begin{eqnarray}
 \phi&=&2b\omega z/c+\pi-2\mu,\\
 \Phi&=&b\omega(z_1+z_2)/c+\Phi_0,
\end{eqnarray}
where $\Phi_0$ is a constant.

The annihilation operators of the out-modes $({\hat c}_1, {\hat c}_2)$, 
are related to the annihilation operators of the in-modes (${\hat a}_1$, ${\hat a}_2$) by the following relations~\cite{Caves:1981}:
\begin{eqnarray}
 {\hat c}_1&=&e^{i\Phi} \l[ -i{\rm e}^{-i\mu} {\hat a}_1 {\rm sin}(\phi/2)
 +{\hat a}_2 {\rm cos}(\phi/2) \r],\\
 {\hat c}_2&=&e^{i\Phi} \l[ {\hat a}_1 {\rm cos}(\phi/2)
 -i{\rm e}^{i\mu} {\hat a}_2 {\rm sin}(\phi/2)\r].
\end{eqnarray}
For the squeezed state $| \psi \rangle$ defined in Eq.~(\ref{def:Sqstate}), 
the expectation value of the difference in number of photons emitted by the
interference arms, and its variance are
\begin{eqnarray}
 n_{\rm out} &\equiv& \langle \psi| (\hat{c}_{2}^\dagger\hat{c}_{2}
 -\hat{c}_{1}^\dagger\hat{c}_{1}) |\psi\rangle \nonumber \\
 &=& {\rm cos}\phi \l[|\alpha|^2-{\rm sinh}^2r\r]~~~~~~~\\ 
 (\Delta n_{\rm out})^2 &=& {\rm cos}^2\phi (|\alpha|^2+2{\rm sinh}^2r~{\rm cosh}^2r)
 +{\rm sin}^2\phi \nonumber \\
 &\times& \biggl[{-\rm sinh}r~ {\rm cosh}r 
 \l(\alpha^2e^{i(\theta+2\mu)}+\alpha^{*^2}e^{-i(\theta+2\mu)}\r) \nonumber \\
 &+&|\alpha|^2 {\rm cosh}(2r)+{\rm sinh}^2r\biggr]
\end{eqnarray}
Like in the previous case, choosing the $\alpha$ to be real and
the squeezing angle $\theta$ to be $-2\mu$, we have
\begin{eqnarray}
 (\Delta n_{\rm out})^2 &=& {\rm cos}^2\phi (\alpha^2+2{\rm sinh}^2r~{\rm cosh}^2r)
  \nonumber \\
 &+&{\rm sin}^2\phi (\alpha^2~e^{-2r}+{\rm sinh}^2r)
\end{eqnarray}
{Thus, the difference in the output photon number changes in $z$,
leading to an error in the displacement of the position of the
end mirrors $[(\Delta z)_{\rm pc}]$ due to the photon-count noise is given by:}
\begin{eqnarray}
\label{caves10}
& & (\Delta z)_{\rm pc}=
 \frac{c}{2b\omega} (\alpha^2-{\rm sinh}^2r)^{-1}  \\
 &\times&\biggl[{\rm cot}^2\phi~(\alpha^2
 +2 {\rm cosh}^2r ~{\rm sinh}^2r)
 +\alpha^2 e^{-2r}+{\rm sinh}^2r \biggr]^{1/2}  \nonumber 
\end{eqnarray}
From the above, one can extract the Caves' limit, i.e.,
${\rm cos}\,\phi=0$, and taking $|\alpha|\gg |{\rm sinh}r|$, to obtain
\begin{equation}
 (\Delta z)_{\rm pc}\approx (c/2b\omega) \alpha^{-1} e^{-r}~,
\end{equation}
although we emphasize that
we will use the exact expression
(\ref{caves10}) to compare with its polymer counterpart in the following section. 

In the next section, we obtain the error in the displacement of the end
mirrors due to the radiation-pressure and photon-count for the polymer
quantization for the advanced LIGO configuration.
\section{Quantum-mechanical noises in advanced LIGO: Polymer Quantization}
\label{sec:Noise2}
To obtain the effects of polymer quantization on fluctuations in
the radiation pressure on the end mirrors (radiation-pressure noise),
and fluctuations in the number of output photons (photon-count noise),
we need to polymer quantize the electromagnetic field in the interferometer
arms.

In the following subsection, we perform polymer quantization of the
electromagnetic field, for small values of $\beta$, and use the approximate polymer creation and
annihilation operators to obtain the quantum-mechanical noises in
the advanced LIGO configuration. 

\subsection{Polymer quantization of electromagnetic field}

Assume that the electromagnetic field is confined in a cavity 
of volume $V$, with periodic boundary conditions. For simplicity, we
assume that the cavity is a cube of length $L$. For small values of $\beta$, one can approximately decompose the electromagnetic
field in the Fourier domain, the Hamiltonian corresponding
to a mode ${\bf k}$ can be written as~\cite{Milonni:1994xx}
\begin{equation}
\label{eq:EMHamiltonian1}
 H_{\bf k}=\frac{1}{2} \int \d V \l(\epsilon_0 E_{\bf k}^{2}
 +\mu_0^{-1} B_{\bf k}^{2}\r).
\end{equation}
Decomposing the vector potential ${\bf A}$ into plane waves in the
Coulomb gauge, electric and magnetic fields are given by~\cite{Milonni:1994xx}:
\begin{eqnarray}
 {\bf E}&=&\frac{1}{\sqrt{\epsilon_0 V}} \sum_{\bf k}
 {\cal E}_{\small {\bf k}} \biggl[\omega_{\small {\bf k}} q_{\small {\bf k}}~{\rm sin}
 (\omega_{\bf k}t-{\bf k}\cdot{\bf r}) \nonumber \\
 \label{def:E}
 &-&p_{\small {\bf k}}~{\rm cos}
 (\omega_{\small {\bf k}}t-{\bf k}\cdot{\bf r})\biggr], \\
 \nonumber \\
 {\bf B}&=&\sqrt{\frac{\mu_0}{V}} \sum_{\bf k}
 ({\hat {\bf k}}\times {\cal E}_{\small {\bf k}}) 
 \biggl[\omega_{\small {\bf k}} q_{\small {\bf k}}~{\rm sin}
 (\omega_{\bf k}t-{\bf k}\cdot{\bf r}) \nonumber \\
 \label{def:B}
 &-&p_{\small {\bf k}}~{\rm cos}
 (\omega_{\small {\bf k}}t-{\bf k}\cdot{\bf r})\biggr],
\end{eqnarray}
where
\begin{eqnarray}
 q_{\small {\bf k}}= \sqrt{\frac{\hbar}{2\omega_{\small {\bf k}}}}
 (a^*_{\small {\bf k}}+a_{\small {\bf k}});~
 p_{\small {\bf k}}=\sqrt{\frac{\hbar \omega_{\small {\bf k}}}{2}}
 (a^*_{\small {\bf k}}-a_{\small {\bf k}}).
\end{eqnarray}
Substituting Eqs. (\ref{def:E}) and (\ref{def:B}) in the
Hamiltonian, i.e., in Eq.~(\ref{eq:EMHamiltonian1}), we get:
\begin{equation}
 H_{\bf k}=\frac{1}{2} \left(p_{\small {\bf k}}^2
 +\omega_{\small {\bf k}}^2 q_{\small {\bf k}}^2\right).
\end{equation}
To proceed with the polymer quantization of electromagnetic fields, for small values of $\beta$, 
we need to polymer quantize the individual harmonic oscillators 
corresponding to different frequency modes ($\omega_{\bf k}$), i. e.,
\begin{equation}
 H_{\bf k(\lambda)}=\frac{1}{2} \l[\l(P_{\small {\bf k}(\lambda)}\r)^2
 +\omega_{\small {\bf k}}^2 q_{\small {\bf k}}^2\r],
\end{equation}
where
\begin{equation}
 P_{\small {\bf k}(\lambda)}=\frac{{\hat U}_{\small {\bf k}(\lambda)}
 -{\hat U}^{\dagger}_{\small {\bf k}(\lambda)}}{2i\lambda},
\end{equation}
with ${\hat U}_{\small {\bf k}(\lambda)}$ as the translation operator
associated with the harmonic oscillator corresponding to the
mode ${\bf k}$ as defined in Eq. (\ref{polymer-operators}). 

As seen in the previous section, to obtain the expressions for
radiation-pressure and photon-count noises, we need to define the
creation and annihilation operators in the Fock space. For the case of
polymer harmonic oscillator, as shown in
Sec.  (\ref{sec:Polymer2}), it is possible to define these operators
in the limit $\beta \ll 1$.
In this limit we can define an approximate polymer
harmonic oscillator state $|{\widetilde \Psi}\rangle$
in the Fock basis (Eq. (\ref{eq:PolyPsiApprox})). 
The approximate state allows us to define the displacement and squeezing
operators in the polymer quantization for the approximate ground state
$| \tilde{0} \rangle$. Though the approximate creation/annihilation
operators satisfy $[{\hat A}_\lambda,{\hat A}^\dagger_\lambda]=1$,
the effect of polymer
quantization effectively comes from the approximated polymer quantum
state $|{\widetilde \Psi}\rangle$. In the rest of this section, we obtain
the fluctuations in the radiation pressure on the mirrors
(radiation-pressure noise) and fluctuations in the number of
output photons (photon-count noise) for the polymer quantized
electromagnetic field in the interferometer arms.

\subsection{Radiation-pressure noise}

Let ${\hat A}_{\lambda(1)}$, ${\hat A}_{\lambda(2)}$ be the polymer annihilation operators corresponding to the input ports ($1^+$) and ($2^+$),
and ${\hat B}_{\lambda(1)}$, ${\hat B}_{\lambda(2)}$ be the polymer
annihilation operators corresponding to the output ports
($1^-$) and ($2^-$) of the interferometer.

In the polymer quantization, the difference between momenta transferred to the end mirrors is given by:
\begin{eqnarray}
 \hat{\widetilde {\cal P}}&\equiv& (2 \nu \hbar \omega/c) \l({\hat B}_{\lambda(2)}^\dagger
 {\hat B}_{\lambda(2)}-{\hat B}_{\lambda(2)}^\dagger
 {\hat B}_{\lambda(2)}\r),\\
 &=&-(2 \nu \hbar\omega/c) \l(e^{i\mu} {\hat A}_{\lambda(1)}^\dagger{\hat A}_{\lambda(2)}
 +e^{-i\mu} {\hat A}_{\lambda(2)}^\dagger{\hat A}_{\lambda(1)}\r).~~~~~
\end{eqnarray}
Here again, we consider squeezing the approximate ground state as shown below,
\begin{equation}
\label{def:PolySqState}
|{\widetilde \psi}\rangle
={\widetilde S}_2(\xi){\widetilde D}_1(\alpha) |\tilde{0} \rangle, 
\end{equation}
where the polymer modified squeezing and displacement operators are
defined as
\begin{eqnarray}
 {\widetilde S}(\xi) &\equiv& {\rm exp}\l\{[\xi^* ({\hat A}_\lambda)^2
 -\xi ~({{\hat A}^\dagger}_\lambda)^2]/2\r\}, \\
 {\widetilde D}(\alpha) &\equiv& {\rm exp}\l(\alpha {{\hat A}^\dagger}_\lambda 
 -\alpha^* {\hat A}_\lambda\r); \,~~ \xi=-r e^{i\theta} \, .
\end{eqnarray}
For the polymer modified squeezed state, we get
\begin{eqnarray}
\label{eq:PolyRP01}
& &  \langle {\widetilde \psi} | {\widetilde {\cal P}} |{\widetilde \psi}\rangle =0,\\
& & \langle {\widetilde \psi} | \l( \Delta{{\widetilde {\cal P}}} \r)^2 |{\widetilde \psi}\rangle=
 (2\, n \, \hbar\omega/c)^2 \biggl[|\alpha|^2 {\rm cosh}(2r)+{\rm sinh}^2r \nonumber \\
 & & ~~~~ +\l(\alpha^2e^{i(\theta+2\mu)}+\alpha^{*^2}e^{-i(\theta+2\mu)}\r) {\rm sinh}r~{\rm cosh}r\biggr] \nonumber \\
 \label{eq:PolyRP02}
 & & ~~~~ \times \frac{\sqrt{\pi}}{\beta}e^{-1/2\beta^2} I_0(1/2\beta^2),\\
\nonumber
\end{eqnarray}
where, $I_0$ is the modified Bessel function of the first kind, 
\begin{equation}
\label{def:beta}
\beta = \sqrt{\frac{\hbar\omega}{M_*c^2}} \, ,
\end{equation}
and $M_*$ be the energy scale (inverse of polymer length scale
$\lambda$) associated with the polymer quantization.

This is an important result, and we would like to stress the
following points: First, like in canonical quantization,
Eq. (\ref{eq:PolyRP01}) implies that the average number of polymer
particles are same in both the interferometer arms. This is not the
case for other quantum gravity inspired models such as GUP. 

\begin{figure}[htb]
\begin{center}
\includegraphics[width=8.4cm]{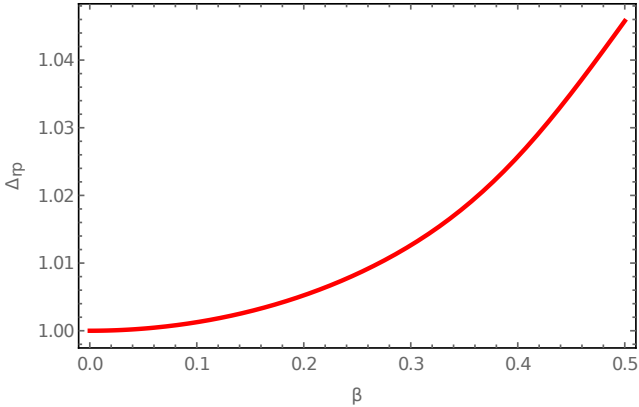} 
\caption{The plot of $\Delta_{\rm rp}$ as a function of $\beta$.}
\label{fig:comparison}
\end{center}
\end{figure}

In the case of GUP, it was shown that the expectation value of
the difference in momentum transfer is non-zero~\cite{Bosso:2018},
however the expectation of difference in momentum transfer vanishes
for the case of polymer quantization. Second, fluctuations in
the momentum transfer, Eq.~(\ref{eq:PolyRP02}),  is different from
that of the
canonical quantization. As in the case of canonical quantization,
choosing the squeezing angle to be $\theta=-2\mu$,  we get
\begin{eqnarray}
 \langle {\widetilde \psi} | \l( \Delta{{\widetilde {\cal P}}} \r)^2 |{\widetilde \psi}\rangle&=&
 \left(\frac{2 \, n \, \hbar\omega}{c}\right)^2 (\alpha^2 {\rm e}^{2r}+{\rm sinh}^2r) \nonumber \\
&\times& \frac{\sqrt{\pi}}{\beta}e^{-1/2\beta^2} I_0(1/2\beta^2),
\end{eqnarray}
Third, the difference in momentum transfer on the end mirrors, for a duration
of time $\tau$, leads to the following error  $[(\Delta {\widetilde z})_{\rm rp}]$ 
\begin{eqnarray}
\label{eq:PolyRP}
 \Delta {\widetilde z}_{\rm rp}=\frac{n \hbar\omega\tau}{mc}
 \l(\alpha^2 {\rm e}^{2r}+{\rm sinh}^2r\r)^{1/2}
 \l[\frac{\sqrt{\pi}}{\beta} \frac{I_0(1/2\beta^2)}{e^{1/2\beta^2}}\r]^{1/2}
\end{eqnarray}
Let $\Delta_{\rm rp}$ be the ratio of 
the radiation-pressure error in the polymer quantization
[$\Delta {\widetilde z}_{\rm rp}$] and the same in the canonical
quantization [$\Delta {z}_{\rm rp}$], i. e.,
\begin{equation}
\Delta_{\rm rp}(\beta) =  \l[\frac{\sqrt{\pi}}{\beta}e^{-1/2\beta^2} I_0(1/2\beta^2)\r]^{1/2},  
\end{equation}
Fig. (\ref{fig:comparison}) contains plot of $\Delta_{\rm rp}(\beta)$
as a function of $\beta$. It is clear that in the limit $\beta \to 0$, the ratio is unity. 
{However, even if $\beta \simeq 0.1$, the difference between the two
quantization schemes is order of {$10^{-3}$}.}
%
\subsection{Photon-count noise}
Let ${\hat C}_{\lambda(1)}$ and ${\hat C}_{\lambda(2)}$
be the annihilation operators corresponding to the out-modes. The annihilation operators
corresponding to the ``in" modes (${\hat A}_{\lambda(1)}$, ${\hat A}_{\lambda(2)}$) are
related to that of ``out" modes by the following relations:
\begin{eqnarray}
 {\widehat C}_{\lambda(1)}&=&e^{i\Phi} \l[{\widehat A}_{\lambda(2)} {\rm cos}(\phi/2)
 -i{\rm e}^{-i\mu} {\widehat A}_{\lambda(1)} {\rm sin}(\phi/2)\r],~~~~~\\
 {\widehat C}_{\lambda(2)}&=&e^{i\Phi} \l[ {\widehat A}_{\lambda(1)} {\rm cos}(\phi/2)
 -i{\rm e}^{i\mu} {\widehat A}_{\lambda(2)} {\rm sin}(\phi/2)\r].
\end{eqnarray}
For the squeezed state $|{\widetilde \psi}\rangle$ defined in Eq. (\ref{def:PolySqState}), 
the expectation value of the difference in the number of polymer photons emitted by the interferometer arms, and its variance are
\begin{eqnarray}
 {\widetilde n}_{\rm out} &\equiv& \langle {\widetilde \psi}| ({{\widehat C}_{\lambda(2)}}^\dagger
 {\widehat C}_{\lambda(2)}
 -{{\widehat C}_{\lambda(1)}}^\dagger{\widehat C}_{\lambda(1)}) |{\widetilde \psi}\rangle \nonumber \\
 &=&{\rm cos}\phi \l(|\alpha|^2-{\rm sinh}^2r\r) \nonumber \\
 \label{eq:PolyPC01}
 &\times& \frac{\sqrt{\pi}}{\beta}e^{-1/2\beta^2} I_0(1/2\beta^2)\\ 
 (\Delta {\widetilde n}_{\rm out})^2 &=& \biggl\{ {\rm cos}^2\phi (|\alpha|^2+2{\rm sinh}^2r~{\rm cosh}^2r)
 +{\rm sin}^2\phi \nonumber \\
 &\times& \biggl[{-\rm sinh}r~ {\rm cosh}r 
 \l(\alpha^2e^{i(\theta+2\mu)}+\alpha^{*^2}e^{-i(\theta+2\mu)}\r) \nonumber \\
 &+&|\alpha|^2 {\rm cosh}(2r)+{\rm sinh}^2r\biggr]+{\rm cos}^2\phi~ {\rm sinh}^4r 
 \nonumber \\
&\times&\l(1-\frac{\sqrt{\pi}}{\beta}e^{-1/2\beta^2} I_0(1/2\beta^2)\r)\biggr\} \nonumber \\
\label{eq:PolyPC02}
&\times & \frac{\sqrt{\pi}}{\beta}e^{-1/2\beta^2} I_0(1/2\beta^2)
\end{eqnarray}
As in the case of canonical quantization, taking $\alpha$ to be real,
and $\theta=-2\mu$ in Eq. (\ref{eq:PolyPC02}), we get:
\begin{eqnarray}
(\Delta {\widetilde n}_{\rm out})^2
&=&\biggl[{\rm cos}^2\phi (\alpha^2+2{\rm sinh}^2r {\rm cosh}^2r)  \nonumber \\
&+&{\rm sin}^2\phi (\alpha^2 e^{-2r}+{\rm sinh}^2r)+{\rm cos}^2\phi~ {\rm sinh}^4r 
 \nonumber \\
&\times&\l(1-\frac{\sqrt{\pi}}{\beta}e^{-1/2\beta^2} I_0(1/2\beta^2)\r) \biggr] \nonumber \\
&\times & \frac{\sqrt{\pi}}{\beta}e^{-1/2\beta^2} I_0(1/2\beta^2)
\end{eqnarray}
The difference in the output photon number changes with respect to $z$,
and hence leads to error in the displacement of the position of the end
mirrors $[(\Delta {\widetilde z})_{\rm pc}]$. Therefore, the photon-count noise is given by:
\begin{eqnarray}
 \Delta {\widetilde z}_{\rm pc}&=&
 \frac{c}{2b\omega} (\alpha^2-{\rm sinh}^2r)^{-1}  \nonumber\\
 &\times&\biggl[{\rm cos}^2\phi (\alpha^2+2{\rm sinh}^2r {\rm cosh}^2r)  \nonumber \\
&+&{\rm sin}^2\phi (\alpha^2 e^{-2r}+{\rm sinh}^2r)+{\rm cos}^2\phi~ {\rm sinh}^4r 
 \nonumber \\
&\times&\l(1-\frac{\sqrt{\pi}}{\beta}e^{-1/2\beta^2} I_0(1/2\beta^2)\r) \biggr]^{1/2} \nonumber \\
 &\times& \l[\frac{\sqrt{\pi}}{\beta}e^{-1/2\beta^2} I_0(1/2\beta^2)\r]^{1/2}
\end{eqnarray}
Again, similar to canonical quantization, setting ${\rm cos}\,\phi=0$, we get 
\begin{eqnarray}
 \Delta {\widetilde z}_{\rm pc}&=&
 \frac{c}{2b\omega} (\alpha^2-{\rm sinh}^2r)^{-1}  \nonumber\\
 &\times&\biggl({\rm sin}^2\phi (\alpha^2 e^{-2r}+{\rm sinh}^2r) \biggr)^{1/2} \nonumber \\
 &\times& \l[\frac{\sqrt{\pi}}{\beta}e^{-1/2\beta^2} I_0(1/2\beta^2)\r]^{1/2}.
\end{eqnarray}
This is an important result and we would like to stress the following points:
First, the fluctuations in the number of output photons in the two
quantization schemes (canonical and polymer) are different. 
Second, let us define $\Delta_{\rm pc}$ as
the ratio of photon-count error in the polymer quantization
[$\Delta {\widetilde z}_{\rm pc}$] and the same in the canonical
quantization [$\Delta {z}_{\rm pc}$]. The ratio $\Delta_{\rm pc}$
is plotted as shown below:
\begin{equation}
\Delta_{\rm pc}(\beta) =  \l[\frac{\sqrt{\pi}}{\beta}e^{-1/2\beta^2} I_0(1/2\beta^2)\r]^{1/2},  
\end{equation}
Thus, $\Delta_{\rm pc}(\beta)$ is identical to 
$\Delta_{\rm rp}(\beta)$, i.e.
the functional dependence of the quantum noises on $\beta$ is identical.
Third, the effects due to Polymer quantization is different from that
of the GUP~\cite{Bosso:2018}. As mentioned earlier, the expectation value of the difference in momentum transfer is
non-zero for the case of GUP~\cite{Bosso:2018}, however, the expectation of difference
in momentum transfer vanishes for the case of polymer quantization. 
On the other hand, the GUP corrections to the radiation-pressure and
photon-count noises are not the same~\cite{Bosso:2018}.
It is also interesting to note that the effects of polymer quantization
on the radiation-pressure and photon-count noises are strikingly different
from that of GUP. 

\section{Conclusions and Discussions}
\label{sec:Conclusions}
We have investigated, in detail, the experimental signatures of the
polymer quantization on the quantum-mechanical noises in the advanced
gravitational wave detectors. This is feasible only if the polymer
quantized electromagnetic modes can be expressed in the Fock space.
We explicitly showed that it is possible to obtain a set of approximate annihilation and
creation operators in the polymer quantized harmonic oscillator in
the limit $\beta \ll 1$.

We used the advanced LIGO configuration to obtain the fluctuations
in the radiation pressure on the mirrors and the fluctuations in the
number of output photons in the polymer quantization
scheme.  The photon-count error ratio
($\Delta_{\rm pc}(\beta)$)  is shown to be identical to the radiation-pressure error ratio
($\Delta_{\rm rp}(\beta)$), where $\beta$ depends on the Polymer scale
$M_*$ and the frequency of the electromagnetic field $\omega$.
Note that, for the case of GUP, it was shown that the error ratios
($\Delta_{\rm rp}$ and $\Delta_{\rm pc}$)
are not identical~\cite{Bosso:2018}.

If the polymer energy scale $M_*$ is assumed to be of the order of Planck scale,
then, for more realistic value of $\omega$, i.e., $2.82\times 10^{14}$ Hz \cite{LIGO-a,LIGO-b},
the parameter $\beta$ is of the order of $10^{-13}$. For small values of $\beta$ the error ratios,
both $\Delta_{\rm rp}$ and $\Delta_{\rm pc}$ (denoted as $\Delta$), can be expanded as
\begin{equation}
\Delta=1+\frac{\beta^2}{8}+O(\beta^4) \, .
\end{equation}
Hence, for realistic values of the frequency of the optical field, the next to leading order
contribution is roughly $10^{26}$ times smaller than the zeroth order contribution.

As mentioned before, we did not take into consideration the effects of modified dispersion 
relation introduced by polymer quantization. Motivated by the modified dispersion relation due 
to polymer quantization of scalar field in Ref.~\cite{Hossain:2010eb}, it possible that the polymer
quantization of Maxwell field can lead to a modified dispersion of the form
\begin{equation}
    \omega^2=|{\bf k}|^2 \l[1+\delta~\beta^2+O(\beta^4)\r],
\end{equation}
in the limit $\beta \ll 1$, where $\delta$ is a constant. If $\delta$ is positive, the
dispersion relation is superluminal, and if it is negative, then the relation is subluminal.
For the above modification, repeating the analysis in Sec. (\ref{sec:Noise2}), the error
ratios $\Delta_{rp}$ and $\Delta_{pc}$ are given by:
\begin{eqnarray}
    \Delta_{rp}&=&1+\l(\frac{1}{8}+\frac{\delta}{2}\r)\beta^2+O(\beta^4)\\
    \Delta_{pc}&=&1+\l(\frac{1}{8}-\frac{\delta}{2}\r)\beta^2+O(\beta^4)
\end{eqnarray}
Note that even if $|\delta| = 1/4$ either one of $\Delta_{rp}$ and $\Delta_{pc}$ will
be non-zero. Therefore, it is evident that the corrections due to modified dispersion
relation is of the same order that we have considered in this work.

In the case of GUP, the expectation value of the difference in momentum transfer is non-zero 
($\langle {\widetilde {\cal P}} \rangle \neq 0$) 
and the quantum noises in the interferometer is lower than the canonical quantization~\cite{Bosso:2018}. However, in the case of Polymer quantization, 
the expectation value of the difference in momentum transfer is zero, and the quantum noises in the interferometer are higher than the canonical quantization. Since, 
$\langle (\Delta {\widetilde {\cal P}})^2 \rangle =  \langle {\widetilde {\cal P}}^2 \rangle
- \langle {\widetilde {\cal P}} \rangle^2$, it is clear that the models that lead to zero (or non-zero) expectation value of the difference in momentum transfer will lead to higher (or lower) quantum noises in the interferometer than the canonical quantization. This feature provides a robust test to distinguish between the two broad categories of quantum gravity phenomenological models.

The analysis in this work is done for a fixed frequency of the
electromagnetic field assuming that the mode decouple and have
a linear-dispersion relation. While it is
true for canonical quantization, it is unclear whether this
feature holds for polymer quantization~\cite{Husain.Louka,Kajuri:2015oza,Stargen:2017}. While the ladder operators
in standard harmonic oscillator are linear combinations of
momentum and position operators,
the approximate ladder operators for the case of approximate
energy eigenfunctions of polymer harmonic oscillator have a
nontrivial combination of ${\hat x}$ and ${\hat p}$:

\begin{eqnarray}
 {\hat A}_{\lambda} &=& \frac{1}{(2\hbar m\omega)^{1/2}}
 \l({\hat P}_\lambda-i\frac{m\omega {\hat x}}{{\rm cos}(\lambda p/\hbar)}\r) \, , \\
 {\hat A}^\dagger_{\lambda}&=&\frac{1}{(2\hbar m\omega)^{1/2}}
 \l({\hat P}_\lambda+i\frac{m\omega {\hat x}}{{\rm cos}(\lambda p/\hbar)}\r).
\end{eqnarray}
The next step in the analysis is to investigate the quantum
mechanical noises due to polymer quantization in the gravitational
wave frequency band for the ground-based and space-based detectors
i.e. $10^{-5}~{\rm Hz}$ to $50~{\rm Hz}$. 
Such an analysis will provide us the tools for testing these results. This work is in progress. 

As we were finalizing this manuscript, the article 
\cite{Gianluca:2019} appeared that discusses plausible 
quantum gravity signatures in future gravitational wave
observations, such as gravitational wave luminosity distance,
the time dependence of effective Planck mass, and also the instrumental
strain noise of interferometers. The focus of this work is different from Ref.
\cite{Gianluca:2019}, which is to analyze the difference in effects due to the
quantization schemes on the interferometer noises.

\begin{acknowledgments}
The work was supported by the Indo-Canadian Shastri Research grant and the Natural
Sciences and Engineering Research Council of Canada. DJS was partially supported by
the DST-Max Planck Partner Group grant, IRCC Seed Grant (IIT Bombay), and the Max
Planck Partner Group "Quantum Black Holes" between CMI Chennai and AEI Potsdam.
SS is partially supported by Homi Bhabha Fellowship Council. DJS and SS thank
Department of Physics and Astronomy, University of Lethbridge for hospitality.
We thank Pasquale Bosso for comments on the earlier draft of the manuscript
\end{acknowledgments}
\appendix

\begin{widetext}
\section{Asymptotic expansion of polymer energy eigenfunctions for large values
of $q$}
\label{Appendix-A}
In this appendix, following Ref. \cite{Blanch:1960}, we explicitly show the
asymptotic expansion of energy eigenfunctions of polymer harmonic oscillator
, $\Psi_{2n}(p)$ and $\Psi_{2n+1}(p)$, for large values of $q$.

Following Eqs.~(\ref{eq:PolyPsi1}) and (\ref{eq:PolyPsi2}), the polymer
energy eigenfunctions in momentum basis are given by:
\begin{eqnarray} 
\label{eq:app-Ce}
\Psi_{2n}(z) &=& \frac{\l(2\beta/\pi\r)^{1/2}}{\l(\hbar m\omega\r)^{1/4}}
 {\rm Ce}_{n}\l(q,z\r),\\
 \label{eq:app-Se}
 \Psi_{2n+1}(z)&=& \frac{\l(2\beta/\pi\r)^{1/2}}{\l(\hbar m\omega\r)^{1/4}}
 {\rm Se}_{n+1}\l(q,z\r),~~
\end{eqnarray}
where $q=1/(4\beta^4)$, and ${\rm Ce}_{n}$, ${\rm Se}_{n}$
are Mathieu functions.

For large values of $q$, the Mathieu functions ${\rm Ce}_n(q,z)$
and ${\rm Se}_{n+1}(q,z)$ can be written as
\cite{Blanch:1960}:
\begin{eqnarray}
\label{eq:app-CeA}
 {\rm Ce}_n(q,z)&=&\frac{(\pi/4)^{1/4} q^{1/8}}{\sqrt{n!}}~U_0
 \l[{\cal Z}_{0,n}(\gamma)+{\cal Z}_{1,n}(\gamma)\r],\\
\label{eq:app-SeA}
 {\rm Se}_{n+1}(q,z)&=&\frac{(\pi/4)^{1/4} q^{1/8}}{\sqrt{n!}}~V_0
 \l[{\cal Z}_{0,n}(\gamma)-{\cal Z}_{1,n}(\gamma)\r] {\rm sin}z,
\end{eqnarray}
where
\begin{eqnarray}
\label{eq:app-Ce1}
 \gamma &=& 2q^{1/4} {\rm cos}z, \\
 U_0&=&1-\frac{(2n+1)}{16\sqrt{q}}+O(1/q),\\
 V_0&=&1+\frac{(2n+1)}{16\sqrt{q}}+O(1/q),\\
 {\cal Z}_{0,n}(\gamma) &=& D_n+\frac{1}{4\sqrt{q}}
 \l(\frac{n!}{16(n-4)!} D_{n-4}(\gamma)-\frac{1}{16}D_{n+4}(\gamma)\r)+O(1/q),\\
 {\cal Z}_{1,n}(\gamma) &=& -\frac{1}{4\sqrt{q}}
 \l(\frac{n(n-1)}{4} D_{n-2}(\gamma)+\frac{1}{4}D_{n+2}(\gamma)\r)+O(1/q),\\
 \label{eq:app-Ce2}
 D_m(\gamma) &\equiv& (-1)^m e^{\gamma^2/4}
 \frac{\d^m}{\d \gamma^m} e^{-\gamma^2/2}=\frac{e^{-\gamma^2/4}}{2^{m/2}}
 H_m(\gamma/\sqrt{2}).
\end{eqnarray}
Substituting Eqs. (\ref{eq:app-Ce1}) - (\ref{eq:app-Ce2}) in
Eqs.~(\ref{eq:app-CeA}) and (\ref{eq:app-SeA}), we obtain
\begin{eqnarray}
\label{eq:app-asymptotic-1}
 {\rm Ce}_n(q,z)=\frac{(\pi/4)^{1/4} q^{1/8}}{\sqrt{n!}}
 \frac{e^{-\gamma^2/4}}{2^{n/2}} \biggl\{{\rm H}_n(\gamma/\sqrt{2})
 &-&\frac{1}{8\sqrt{q}}\biggl[\frac{1}{32}H_{n+4}(\gamma/\sqrt{2})
 +\frac{1}{4}H_{n+2}(\gamma/\sqrt{2})+\frac{2n+1}{2}H_{n}(\gamma/\sqrt{2}) \nonumber\\
 &+&n(n-1)H_{n-2}(\gamma/\sqrt{2})
 -\frac{n!}{2(n-4)!} H_{n-4}(\gamma/\sqrt{2})\biggr]+O(1/q)\biggr\},~~~~
\end{eqnarray}

\begin{eqnarray}
\label{eq:app-asymptotic-2}
 {\rm Se}_{n+1}(q,z)=\frac{(\pi/4)^{1/4} q^{1/8}}{\sqrt{n!}}
 \frac{e^{-\gamma^2/4}}{2^{n/2}} \biggl\{{\rm H}_n(\gamma/\sqrt{2})
 &-&\frac{1}{8\sqrt{q}} \biggl[\frac{1}{32}H_{n+4}(\gamma/\sqrt{2})
 -\frac{1}{4}H_{n+2}(\gamma/\sqrt{2})-\l(\frac{2n+1}{2}-\gamma^2\r)
 H_{n}(\gamma/\sqrt{2}) \nonumber\\
 &-&n(n-1)H_{n-2}(\gamma/\sqrt{2})
 -\frac{n!}{2(n-4)!} H_{n-4}(\gamma/\sqrt{2})\biggr]+O(1/q)\biggr\},
\end{eqnarray}
where $H_n$ is the Hermite polynomial. Making use of the variables defined in Eqs.~(\ref{eq:variable-defs-1}) and (\ref{eq:variable-defs-2}),
and the asymptotic expansions in Eqs. (\ref{eq:app-asymptotic-1})
and (\ref{eq:app-asymptotic-2}), we obtain
\begin{eqnarray}
\label{eq:appendix-asymptotic-1}
 \Psi_{2n}(p)=\l(\frac{1}{\pi\hbar m\omega}\r)^{1/4}
 \frac{e^{-\alpha^2/2}}{2^{n/2}\sqrt{n!}} \biggl\{{\rm H}_n(\alpha)
 &-&\frac{\beta^2}{4}\biggl[\frac{1}{32}H_{n+4}(\alpha)
 +\frac{1}{4}H_{n+2}(\alpha)+\frac{2n+1}{2}H_{n}(\alpha) \nonumber\\
 &+&n(n-1)H_{n-2}(\alpha)
 -12~nC_4 H_{n-4}(\alpha)\biggr]+O(\beta^4)\biggr\},
\end{eqnarray}

\begin{eqnarray}
\label{eq:appendix-asymptotic-2}
 \Psi_{2n+1}(p)=\l(\frac{1}{\pi\hbar m\omega}\r)^{1/4}
 \frac{e^{-\alpha^2/2}}{2^{n/2}\sqrt{n!}} \biggl\{{\rm H}_n(\alpha)
 &-&\frac{\beta^2}{4}\biggl[\frac{1}{32}H_{n+4}(\alpha)
 -\frac{1}{4}H_{n+2}(\alpha)-\l(\frac{2n+1}{2}-2\alpha^2\r)H_{n}(\alpha) \nonumber\\
 &-&n(n-1)H_{n-2}(\alpha)
 -12~nC_4 H_{n-4}(\alpha)\biggr]+O(\beta^4)\biggr\},
\end{eqnarray}
where $\alpha={\rm sin}(\lambda p/\hbar)/\beta$. Note that in the limit
$\lambda \to 0$, one can obtain the energy eigenfunctions of the
canonically quantized simple harmonic oscillator.
\end{widetext}

\end{document}